\documentclass[%
reprint,
superscriptaddress,
 amsmath,amssymb,
 aps,
floatfix,
]{revtex4-2}
\usepackage[main=english,vietnamese]{babel}
\usepackage{comment}
\usepackage{graphicx}
\usepackage{dcolumn}
\usepackage{bm}
\usepackage{xcolor}
\usepackage{ulem}

\DeclareMathAlphabet{\vecfont}{OT1}{cmr}{bx}{it}

\begin{document}

\newcommand{\avsb}{\textit{A}V$_3$Sb$_5$}
\newcommand{\csvsb}{CsV$_3$Sb$_5$}
\newcommand{\kvsb}{KV$_3$Sb$_5$}
\newcommand{\rbvsb}{RbV$_3$Sb$_5$}
\newcommand{\atibi}{ATi$_3$Bi$_5$}
\newcommand{\cscrsb}{CsCr$_3$Sb$_5$}
\newcommand{\avsbsn}{AV$_3$Sb$_{5-x}$Sn$_x$}
\newcommand{\amx}{AM$_6$X$_6$}
\newcommand{\svs}{ScV$_6$Sn$_6$}
\newcommand{\rvs}{\textit{R}V$_6$Sn$_6$}

\preprint{APS/123-QED}

\title{$^{51}$V NMR evidence for interlayer-modulated charge order and a first-order low-temperature transition in \csvsb\ }

\author{Xiaoling Wang} \email[Corresponding author: ]{xiaoling.wang@csueastbay.edu}
\affiliation{Department of Chemistry and Biochemistry, California State University East Bay, Hayward, California 94542, USA}

\author{Arneil P. Reyes}
\affiliation{National High Magnetic Field Laboratory, Tallahassee, Florida 32310, USA}

\author{Hrishit Banerjee}
\affiliation{University of Dundee, Dundee DD1 4HN, Scotland, United Kingdom}

\author{Andrea N. Capa Salinas}
\affiliation{Materials Department, University of California, Santa Barbara, California 93106, USA}

\author{Stephen Wilson}
\affiliation{Materials Department, University of California, Santa Barbara, California 93106, USA}

\author{Brenden R. Ortiz}%
\affiliation{Materials Science and Technology Division, Oak Ridge National Laboratory, Oak Ridge, Tennessee 37831, USA}%

\begin{abstract}

Charge order in the kagome superconductor \csvsb\ exhibits a complex three-dimensional organization and intermediate-temperature anomalies whose bulk character has remained unsettled. We use orientation-dependent $^{51}$V nuclear magnetic resonance (NMR) as a site-selective probe to determine the stacking of the charge density wave (CDW) state and its thermal evolution. Below $T_{\mathrm{CDW}}\!\approx\!94$~K, the field-linear splitting of the $^{51}$V central transition together with the anisotropy of the Knight-shift tensor identify an interlayer-modulated $3\mathbf{q}$ CDW whose local environments are consistent with a four-layer $2\times2\times4$ stacking with mixed trihexagonal/Star-of-David distortions, in agreement with synchrotron x-ray determinations. For comparison, \rbvsb\ serves as a reference exhibiting a uniform trihexagonal $2\times2\times2$ stacking, allowing us to isolate features unique to the $2\times2\times4$ state in \csvsb. With $\mathbf{H}_0\!\parallel\!c$, the $^{51}$V quadrupolar satellites through the intermediate temperature scale near $T_{\mathrm{CO}}\!\approx\!65$~K reorganize into two well-resolved electric-field-gradient manifolds that coexist over a finite interval; their relative spectral weights interchange on cooling while the total integrated satellite intensity remains conserved and $\nu_Q$ within each manifold is nearly temperature independent. The coexistence without critical broadening, together with conserved intensity, provides bulk evidence consistent with a first-order charge-order transition near $T_{\mathrm{CO}}$. Our measurements do not resolve whether this lower-temperature transition corresponds to a distinct in-plane order or a reorganization of the $3\mathbf{q}$ state; rather, they delimit this window and provide bulk, site-resolved constraints that connect prior reported anomalies to a thermodynamic first-order transition.

\end{abstract}

\maketitle



\textit{Introduction.}
The recent discovery of the kagome superconductors \avsb\ (\textit{A} = K, Rb, Cs) has stimulated extensive research, particularly focusing on their complex charge density wave (CDW) states\cite{ortiz2019physrevmater,ortiz2020physrevlett,yin2021chinphyslett}. In these materials, CDW formation is closely tied to Fermi surface nesting and proximity to van Hove singularities (VHSs) near the Fermi level\cite{tan2021physrevlett,kang2022natphys,hu2022natcommun,christensen2021physrevb,labollita2021physrevb}. The CDW phase involves subtle structural distortions within the vanadium sublattice, establishing a three-dimensional ordered state with well-defined phase coherence across the kagome planes~\cite{li2021physrevx,liang2021physrevx}. The distortion is characterized by a $3\mathbf{q}$ breathing mode pattern, forming a trihexagonal (TrH) or a Star of David (SoD) arrangement. Whereas \kvsb\ and \rbvsb\ predominantly exhibit a staggered TrH pattern, \csvsb\ displays a more intricate arrangement~\cite{Ortiz2021,kang2023natmater}. Scanning tunneling microscopy (STM) studies have reported alternating layers of SoD and TrH distortions, highlighting the intricate layered nature of the CDW order~\cite{Zhao2021}. Synchrotron X-ray diffraction reveals that the CDW evolves from intermediate $2\times2\times1$ and $2\times2\times2$ superstructures to a stable $2\times2\times4$ phase at lower temperatures, characterized by specific stacking patterns of SoD and TrH distortions~\cite{kautzsch2023physrevmater}. Raman spectroscopy experiments have identified pronounced phonon anomalies indicative of strong electron-phonon coupling involved in CDW formation~\cite{Liu2022}. Complementary density functional theory (DFT) and angle-resolved photoemission spectroscopy (ARPES) attribute the CDW instabilities primarily to electronic nesting involving vanadium-derived VHS near the $M$ and $L$ points of the Brillouin zone~\cite{tan2021physrevlett,kang2022natphys}. Furthermore, coherent phonon spectroscopy provides direct evidence of simultaneous phonon condensation at these critical wavevectors, underscoring the multi-modal character of the CDW transition~\cite{Ratcliff2021}. 

In this study, we use angle‑resolved $^{51}$V nuclear magnetic resonance (NMR) as a bulk, site‑selective probe to address two open issues. First, while synchrotron X‑ray and STM have established that \csvsb\ often hosts a mixed‑layer CDW with a $2\times2\times4$ modulation distinct from the uniform staggered‑TrH $2\times2\times2$ order in \rbvsb\, a direct, bulk spectroscopic discriminator between these three‑dimensional microstructures has been lacking. Second, although surface‑ and optical‑probe measurements—including Raman/optical phonon anomalies and SI‑STM reports of $4a_0$ “stripe‑like” modulations—have revealed anomalies near $60$ K in \csvsb\, they do not establish a bulk symmetry‑lowering phase transition or determine its thermodynamic order. For comparison, \rbvsb\ serves as a reference exhibiting a uniform TrH $2\times2\times2$ stacking, allowing us to isolate features unique to the $2\times2\times4$ state in \csvsb. We show that the $^{51}$V Knight‑shift tensor cleanly separates the interlayer‑modulated CDW microstructures of \csvsb\ and \rbvsb\ right below $T_{\mathrm{CDW}}$, providing a bulk spectroscopic discriminator consistent with the diffraction‑inferred stacking in each compound. To our knowledge, prior NMR/nuclear quadrupole resonance (NQR) studies did not explicitly delineate this Rb–Cs contrast via a Knight‑shift splitting immediately below $T_{\mathrm{CDW}}$\cite{Mu_2021,song2022,luo2022npjquantummater,Zheng2022Nature,Feng2023npjQM,frassineti2023,Zhang2024APL,Feng2025NatCommun,Oey2022PRMaterials,CapaSalinas2023Frontiers}; earlier evidence for differing stacking largely came from diffraction and STM. At lower temperature in \csvsb\, the $^{51}$V quadrupolar satellites reorganize into two coexisting electric field gradient (EFG) environments, establishing bulk evidence consistent with a first‑order charge‑order transition into a symmetry‑lowered phase. Density functional theory (DFT) calculations reproduce the increased spread of EFG anisotropy and point to enhanced in-plane V-$d$/Sb-$p$ hybridization. While NMR does not determine the in‑plane wave vector, these bulk fingerprints are compatible with the unidirectional ($4a_0$) correlations seen by SI‑STM near this temperature scale.

\section{Experimental Methods}

\subsection{Nuclear Magnetic Resonance (NMR) Methods}

Orientation-dependent $^{51}$V NMR experiments were performed on single-crystal samples of \csvsb, prepared with optimized crystalline quality for accurate spectral measurements, according to\cite{ortiz2020physrevlett}. All NMR measurements used a specialized two-axis rotation device (illustrated in Fig.~S1), essential for resolving subtle electronic and lattice modulations associated with CDW transitions.

Angle-resolved spectra were acquired by rotating the crystal with respect to the static field $\mathbf{H}_0$ using a two-axis goniometer. We define $\phi$ as the azimuthal rotation about the crystallographic $c$ axis within the $ab$ plane and $\theta$ as the polar angle between $\mathbf{H}_0$ and the $c$ axis ($\theta=0^{\circ}$ for $\mathbf{H}_0\!\parallel\!c$, $\theta=90^{\circ}$ for $\mathbf{H}_0\!\parallel\!ab$); see Fig.~S2 and SI~Sec.~I. This rigorous orientation control allowed for the distinct identification and characterization of unique vanadium sites emerging at and below the CDW transition temperature. The detailed analysis involved the extraction of the Knight shift ($K$) and EFG ($V$) tensor components from the angle-dependent spectral patterns. These tensors were determined by exact diagonalization of the Zeeman and quadrupole Hamiltonians for the $^{51}$V nuclei, followed by precise mathematical transformations among the crystal lattice, the rotation device axes, and the tensor coordinates, as illustrated in SI Section I.

\subsection{Density Functional Theory (DFT) Calculations}

Structural relaxations within DFT were performed using a plane-wave basis set and projector-augmented wave potentials (PAW)~\cite{blochl}, as implemented in the Vienna \textit{Ab initio} Simulation Package (VASP)~\cite{kresse, kresse01}.

In all our DFT relaxation calculations, we used the generalized gradient approximation using the Perdew-Burke-Ernzerhof (PBE) exchange correlation functional~\cite{pbe}. Ionic relaxations were performed using the Vienna \textit{Ab initio} Simulation Package (VASP), allowing internal atomic positions to relax until the forces were less than 0.005\,eV/\AA. An energy cutoff of 600\,eV and an $8\times8\times4$ Monkhorst–Pack $k$-point mesh ensured good convergence of the total energy. Computational details regarding the  EFG tensor are included in the SI Section II.

\section{Results and Discussion}

\subsection{Knight shift splitting and implications for the CDW superstructures}
\begin{figure}
\vspace*{0.2cm}
\includegraphics[scale=0.16]{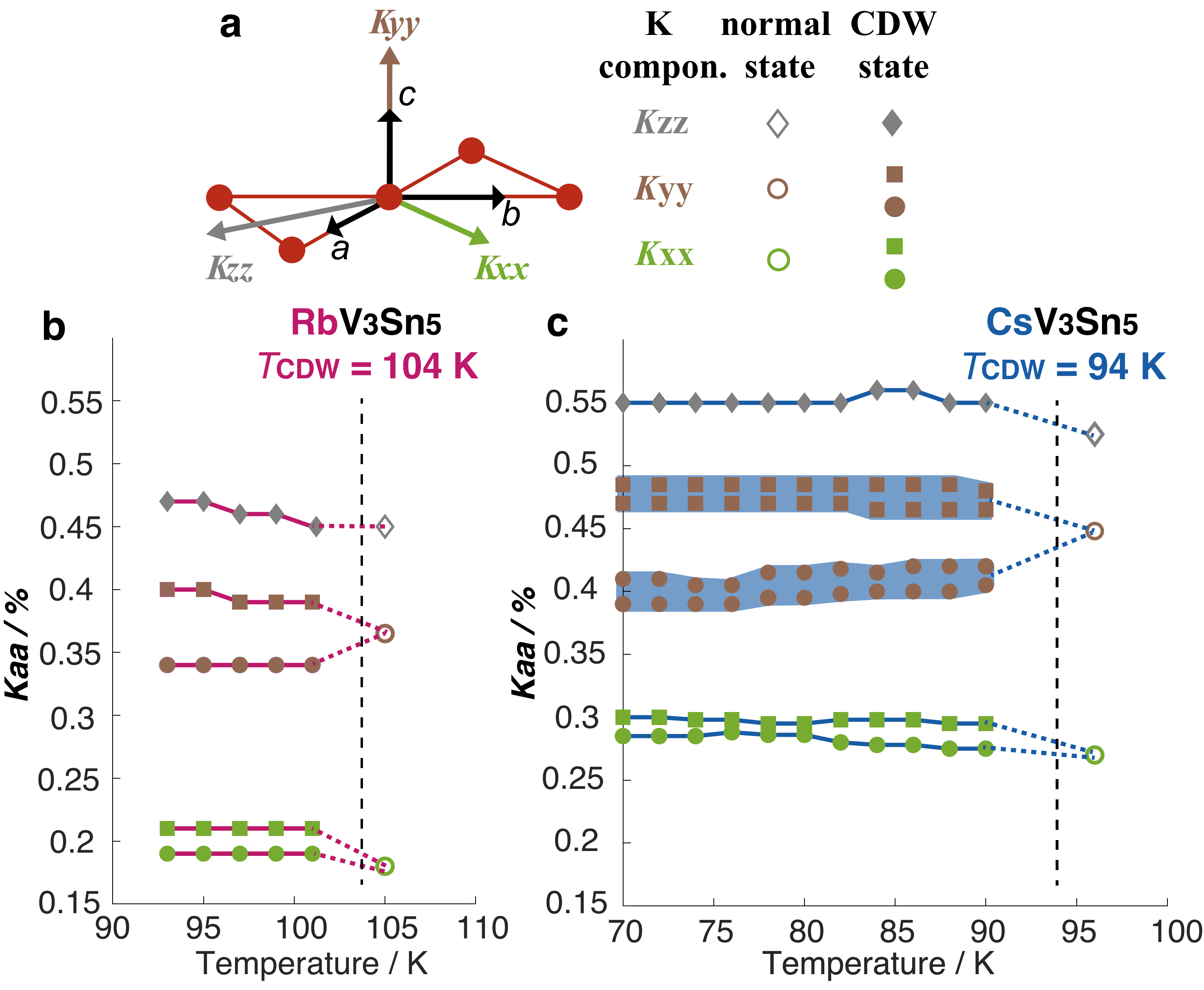}
\caption{\label{fig:epsart}{$^{51}$V Knight-shift tensor through the CDW transition in \rbvsb\ and \csvsb.}
(a) Definition of principal components with respect to crystal axes.Temperature dependence of $K_{xx}$, $K_{yy}$, and $K_{zz}$ in \rbvsb\ across $T_{\mathrm{CDW}}=104$ K (b) and for \csvsb\ across $T_{\mathrm{CDW}}=94$ K (c).
Circles and squares denote the higher- and lower-frequency central-transition components.}
\end{figure}

\begin{figure*}[htbp]
    \centering
\includegraphics[scale=0.17]{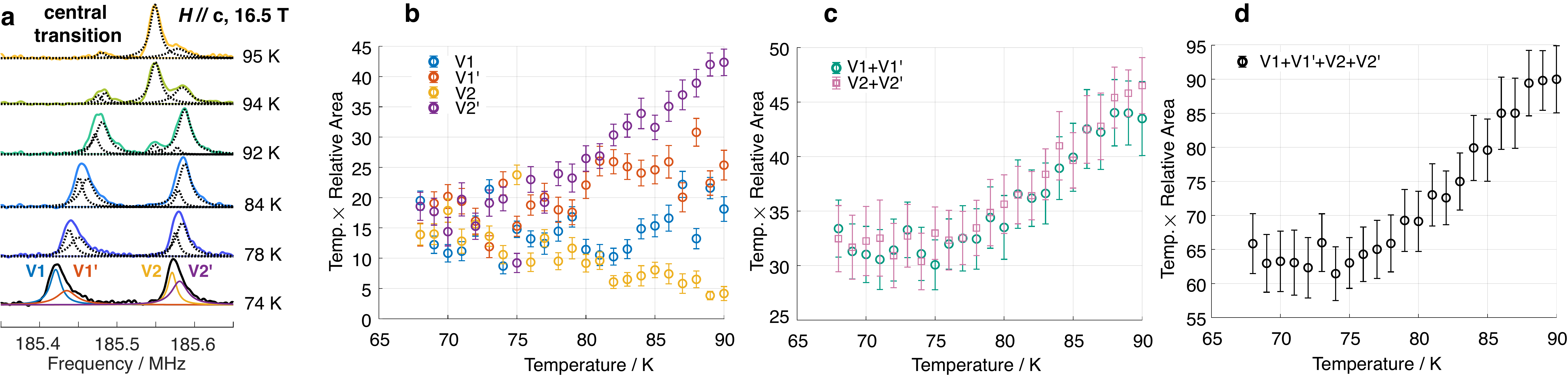}
\caption{\label{fig:epsart} Temperature evolution of the $^{51}$V central transition and component intensities in \csvsb\ ($H_0 \parallel c$). (a) Stacked central-transition spectra measured at 16.5 T for selected temperatures. (b) Boltzmann-corrected integrated areas $I(T)\,T$ for the four fitted components (V1, V1$'$, V2, V2$'$). (c) Pairwise sums $I(T)\,T$ for V1+V1$'$ and V2+V2$'$. (d) Total $I(T)\,T$. Error bars are fit uncertainties.}
\end{figure*}

We performed exact diagonalization of the combined Zeeman and quadrupole Hamiltonians for each distinct vanadium site, fitting angle-dependent $^{51}$V NMR spectral data measured at multiple crystal orientations. Through systematic fitting of these angular dependencies, we determined tensor components for each non-equivalent vanadium site, capturing local spin-density modulations induced by the CDW. The anisotropy of the Knight shift provides a sensitive probe of both the in-plane $2\times 2$ reconstruction and the interlayer phasing of the charge order. 

Specifically, below $T_{\mathrm{CDW}}$, as illustrated in Fig.~1, our analysis clearly revealed that the Knight shift component in the plane ($K_{xx}$), oriented in the kagome plane, splits into two distinct resonances below the CDW transition, reflecting the formation of two crystallographically distinct vanadium sites in both \rbvsb\ and \csvsb. The perpendicular in-plane component ($K_{zz}$) remains unsplit in both the Cs and Rb variants, indicating negligible modulation in spin susceptibility or density of states along that particular direction. A key contrast between the Cs and Rb variant is that the component outside the plane ($K_{yy}$), oriented along the crystal $c$ axis, the low $T$ spectra exhibit two components without further resolvable splitting, consistent with a uniform TrH $2 \times 2 \times 2$ stacking structure\cite{frassineti2023,Wenzel2022PRB,kautzsch2023physrevmater}. However, in \csvsb, by contrast, each vanadium site splits into a doublet (V1:V1$'$ and V2:V2$'$) as illustrated in Fig. 2a, and the field-linear dependence of the central transition splitting identifies a Knight-shift origin rather than second-order quadrupolar effects. The additional modulation of the c axis, often described as a competition among $2 \times 2 \times 2$ and $2 \times 2 \times 4$ stackings with mixed layers of TrH and $\mathrm{SoD}$, has been reported in diffraction and STM studies of \csvsb\ and is absent (within current resolution) in \rbvsb\ and \kvsb \cite{liang2021physrevx,li2021physrevx,Stahl2022PRB,Xiao2023PRR,kautzsch2023physrevmater,Jiang2021NatMater,Kato2022ComMat}. The temperature evolution of Knight shift anisotropy across $T_{\mathrm{CDW}}$ is consistent with a bond-centered, $\mathbf{3q}$ CDW that reconstructs the Fermi surface and opens anisotropic, partially gapped spectra,\cite{tan2021physrevlett,Kato2022ComMat,NakayamaPRB2021} with interlayer phase correlations mediated in part by Sb-derived states.\cite{JeongPRB2022,RitzPRB2023,LiPRB2023}

The temperature evolution of the integrated echo intensity corrected for Boltzmann polarization, $I(T)\,T$, provides a quantitative proxy for the number of nuclei contributing to the echo at fixed echo spacing. For the V1/V1$'$ and V2/V2$'$ families the quantities $I_{V_1}(T)+I_{V_1^\prime}(T)$ and $I_{V_2}(T)+I_{V_2^\prime}(T)$ decrease together upon cooling from $90$~K to $75$~K and then become temperature independent between $75$~K and $68$~K; the total $I_{V_1}(T)+I_{V_1^\prime}(T)+I_{V_2}(T)+I_{V_2^\prime}(T)$ follows the same trend (Figs.~2(c)--2(d)).

Between $90$~K and $75$~K, the product $I(T)\,T$ extracted from two-pulse echoes at fixed echo spacing $2\tau$ decreases even though the repetition time is long compared to $T_1$ and the central-transition $T_2$ exhibits only modest fluctuations within experimental uncertainty, with no systematic decrease. For a half-integer quadrupolar nucleus such as ${}^{51}$V ($I{=}7/2$), the central-transition Hahn-echo amplitude contains oscillatory envelope terms arising from second-order quadrupolar effects; the positions of nodes and antinodes shift with the EFG parameters $\nu_Q$ and $\eta$ and with field orientation.\cite{Man1997PRB,Walker1985PRB} As $\nu_Q$ and $\eta$ evolve across the CDW regime of \csvsb,\cite{luo2022npjquantummater,Mu_2021} a fixed $2\tau$ can move closer to an envelope minimum on cooling, thereby suppressing the echo amplitude although the long-time decay that defines $T_2$ remains essentially unchanged. In our data this short-time oscillatory envelope develops between 90 and 75~K as the CDW forms and then persists with approximately constant amplitude at lower temperatures; the effect is particularly evident for $2\tau\simeq100$--$110~\mu\mathrm{s}$ (SI Fig.~S6). A strong dependence of apparent spectral weight on the measurement time scale $2\tau$ near charge order has been reported in other materials as well.\cite{Imai2017PRB} In addition, central-transition excitation is sensitive to flip angles and RF bandwidth: as $|\nu_Q|$ grows or the lineshape widens, rectangular pulses under-excite portions of the central-transition manifold unless flip angles are re-optimized or broadband/frequency-stepped schemes are used.\cite{ODell2010JMR,Kwak2003JMR} For these reasons we do not interpret absolute $I(T)\,T$ as a volumetric measure in this temperature window. Instead, to quantify the interlayer modulation we analyze spectral-area ratios at fixed \(T\) of the doubled components,
\begin{equation}
R_i(T)=\frac{I_{V_i'}(T)}{I_{V_i}(T)+I_{V_i'}(T)}\quad (i=1,2),
\end{equation}
which are insensitive to global gain and to short-time envelope nodes at fixed $T$ under identical acquisition conditions. The monotonic evolution of \(R_2(T)\) and \(R_1(T)\) evidences a redistribution among the layer-selective V environments that arise only in a \(2\times2\times4\) interlayer modulation, consistent with synchrotron x-ray reports of temperature-driven reorganization/coexistence of \(2\times2\times2\) and \(2\times2\times4\) stackings in \csvsb.

\subsection{Quadrupolar EFG anomalies and first-order charge order at $\sim$65~K}

\begin{figure*}[htbp]
    \centering
\includegraphics[scale=0.42]{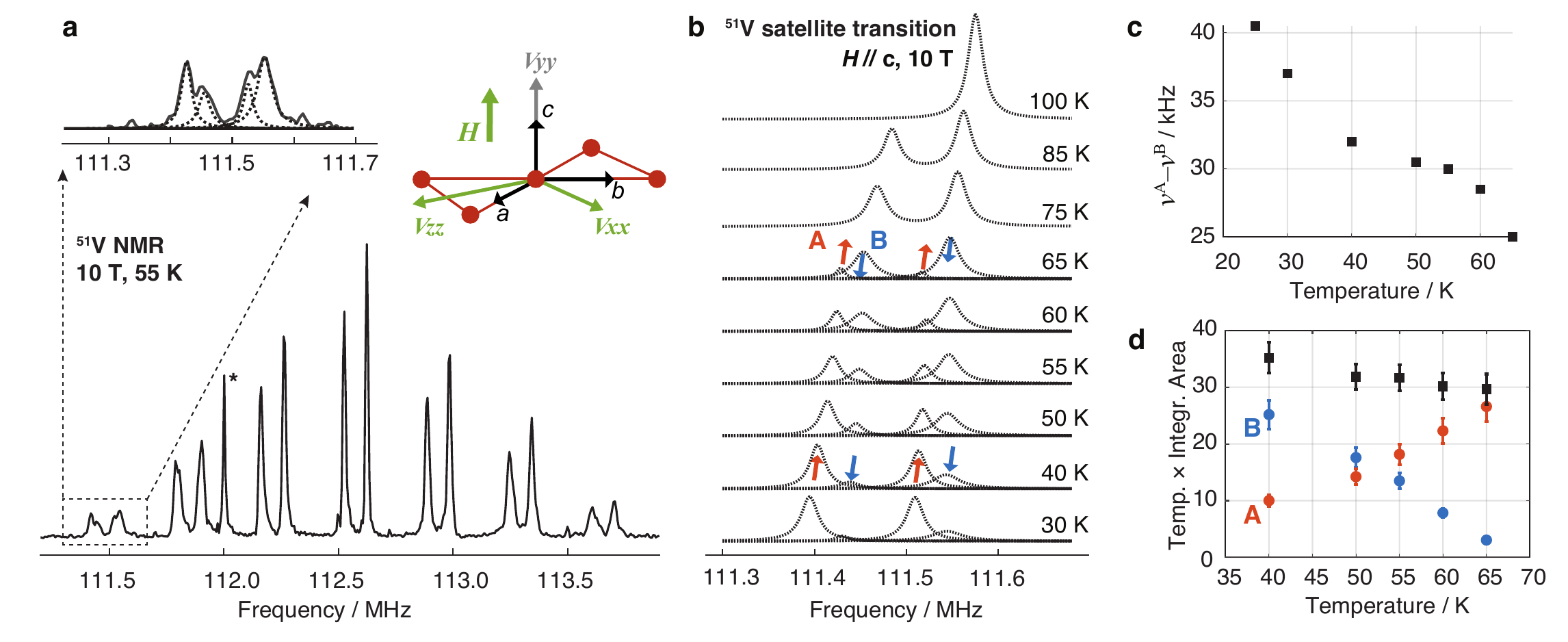}
\caption{\label{fig:epsart} $^{51}$V quadrupolar satellites in \csvsb\ at the intermediate transition $T_{\mathrm{CO}}\!\approx\!65$~K (\,$\mathbf{H}_0 \parallel c$\,) within the $2\times2\times4$ CDW background.
(a) Representative spectrum at 55~K and 10~T; the expanded view highlights fitted satellite peaks for distinct in-plane EFG environments.(b) Temperature series of the satellite transition manifolds showing the emergence of two well-resolved sets on cooling. (c) Temperature dependence of the inter-manifold splitting. (d) Relative spectral weights of the two EFG manifolds; the sum of integrated satellite intensity is conserved within experimental uncertainty. Error bars denote fit uncertainties.}
\end{figure*}

To interrogate charge–lattice modulations in \csvsb\ in the CDW state, we track the temperature evolution of the $^{51}$V quadrupolar satellites with $\mathbf{H}_0 \parallel c$ (Fig.~3a). In this geometry the satellite frequencies are set by the in-plane principal components of the EFG tensor and their orientations with respect to $\mathbf{H}_0$. Fig.~3 summarizes the temperature evolution of the $^{51}$V quadrupolar satellites across the intermediate temperature scale at $T_{\mathrm{CO}}\!\approx\!65$~K. On cooling through $T_{\mathrm{CO}}$ the spectra undergo an abrupt rearrangement: a single set of satellites above $T_{\mathrm{CO}}$ gives way to two well-resolved sets with distinct quadrupole splittings $\nu_Q^{A}$ and $\nu_Q^{B}$, while the total integrated satellite intensity remains conserved within experimental uncertainty. The sudden appearance of two inequivalent EFG environments, the absence of critical broadening, and the persistence of two-component spectra over a finite temperature interval are taken as evidence for a first-order transition at $T_{\mathrm{CO}}$ accompanied by in-plane phase segregation. In this regime the relative spectral weights of the two satellite manifolds evolve with temperature, whereas each $\nu_Q$ is nearly $T$ independent, consistent with coexisting domains with different local EFGs whose volume fractions gradually interchange on cooling. Because two distinct satellite sets appear for $\mathbf H_0\!\parallel\!c$, the data demonstrate two groups of V sites with different EFG tensors. This observation is consistent with a reduction of the average in-plane rotational symmetry within the charge-ordered state, although NMR alone cannot fix the in-plane wave vector nor exclude multi-domain or stacking-registry scenarios.

This intermediate transition is distinct from the primary CDW transition at $T_{\mathrm{CDW}}\!\approx\!94$~K in CsV$_3$Sb$_5$, which is already known to be first order and a 3q ground state with interlayer-shifted tri-hexagonal distortions.\cite{Ratcliff2021} By contrast, prior work on the lower onset ($T\!\sim\!60$--$70$~K) reported lattice/electronic anomalies but did not establish the order of the transition: coherent-phonon spectroscopy detected the emergence of an additional mode near $T^{\ast}\!\approx\!60$~K and argued that this feature ``appears at $T^{\ast}\!\approx\!60$~K, well below $T_{\mathrm{CDW}}$'' and may be related to uniaxial ($1\mathbf{q}$) order, possibly as a crossover or order–disorder phenomenon,\cite{Ratcliff2021} polarization-resolved Raman scattering modeled the anomaly with coupled primary/secondary-like order parameters with $T^{\ast}$ introduced phenomenologically ($\sim\!70$--$80$~K) but did not identify a thermodynamic phase transition\cite{Wu2022PRB,Wulferding2022}, and SI-STM observed $4a_0$ stripe order below $\sim\!60$~K while emphasizing that targeted bulk scattering would be needed to determine whether the stripe order forms a bulk phase.\cite{Zhao2021} Our $^{51}$V NMR data provide bulk, site-resolved evidence for a first-order transition at $T_{\mathrm{CO}}$, thereby resolving this ambiguity. 

\begin{figure}
\vspace*{0.2cm}
\includegraphics[scale=0.32]{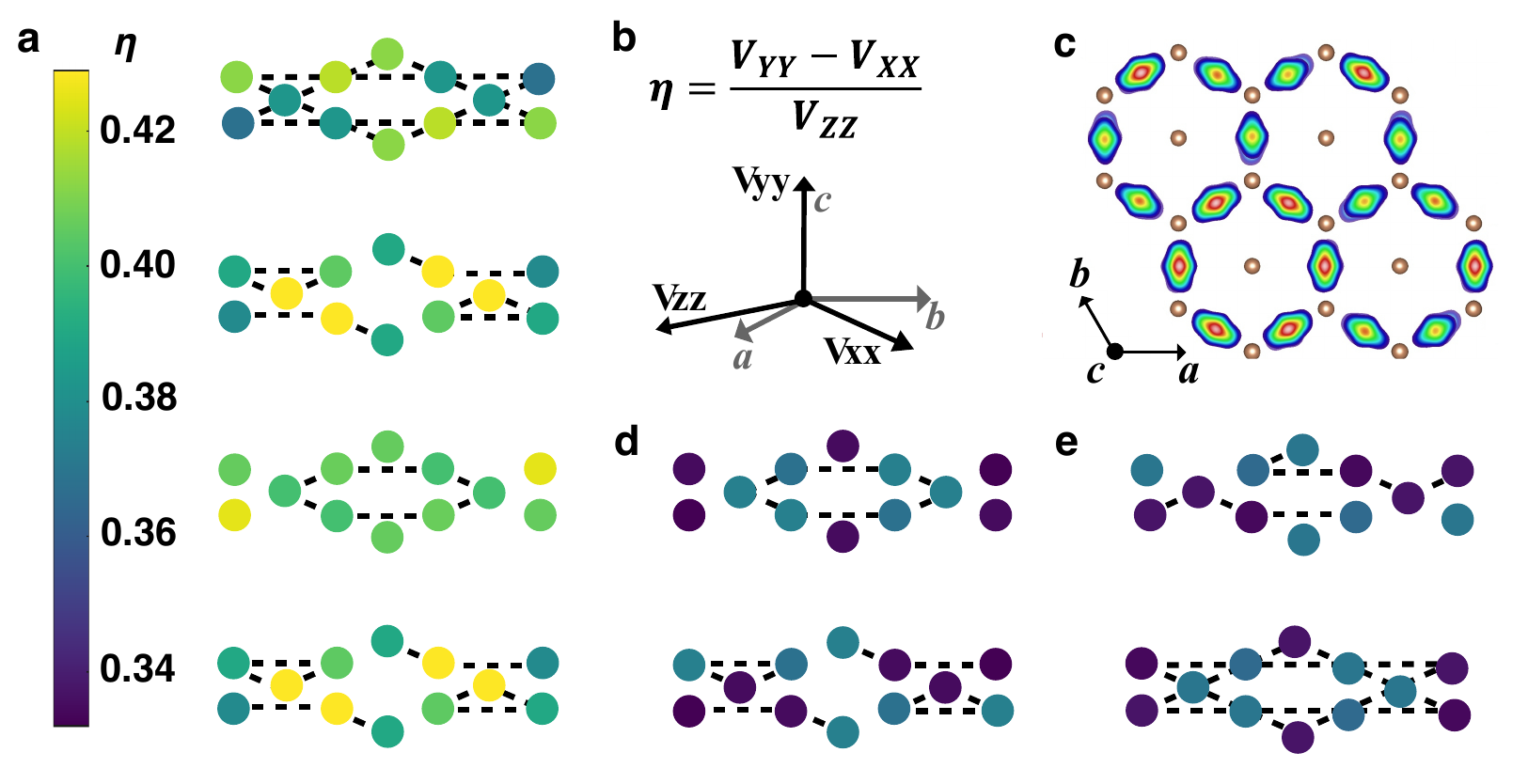}
\caption{\label{fig:epsart} DFT-calculated EFG anisotropy $\eta$ and partial charge density for \csvsb. Calculations use synchrotron-refined structures from Ref.~[14] representative of (low-$T$) 11 K and (high-$T$) 90 K states.
(a) Distribution of the EFG anisotropy $\eta$ over V sites for the 11 K structure (color scale at left).(b) Orientation of principal EFG axes $(V_{xx},V_{yy},V_{zz})$ relative to crystal axes $(a,b,c)$. (c) Partial charge density near $E_F$ for the 11~K structure of \csvsb. Real-space partial charge density associated with states within a narrow energy window around $E_F$ (selected from the DOS peak; see SI for energy window and isovalue), computed for the synchrotron‑refined 11~K structure and viewed along $c$ (projected onto the kagome layer). (d),(e) Distributions of $\eta$ for 90 K structures.}
\end{figure}

In the context of the SI\mbox{-}STM observations\cite{Zhao2021}, we thus attribute the two EFG environments below $T_{\mathrm{CO}}$ to stripe-like charge order that breaks the in-plane sixfold symmetry of the CDW background. For the quadrupolar nucleus ${}^{51}$V ($I{=}7/2$), we parameterize the EFG by principal components $(V_{xx},V_{yy},V_{zz})$ with $|V_{zz}|\!\ge\!|V_{yy}|\!\ge\!|V_{xx}|$ and define $\eta\!\equiv\!(V_{xx}-V_{yy})/V_{zz}$. In our convention informed by the refined structures, $V_{zz}$ and $V_{xx}$ lie approximately in-plane while $V_{yy}\!\approx\!c$, so $\eta$ quantifies the in-plane vs out-of-plane charge anisotropy at V. To relate the NMR signatures to specific structural models, we computed V-site EFGs using the synchrotron-refined structures of \csvsb\ at 90~K and 11~K.\cite{kautzsch2023physrevmater}. The 90~K models, representative of the interlayer-modulated $3\mathbf q$ background above $T_{\mathrm{CO}}$, yields a comparatively narrow spread of $V_{zz}$ and $\eta$ across V sites (Figs.~4d,4e), whereas for the 11~K structure the calculated EFGs segregate into several well-separated clusters with larger $\eta$ contrast (Fig.~4a), mirroring the discrete environments resolved in Fig.~3.

Within these structural models, the orbital-resolved partial charge densities near $E_F$ (Fig.~4c) show a stronger, directional V-$3d$/Sb-$5p$ mixing anisotropy at 11~K than at 90~K. This trend accords with temperature‑dependent X‑ray absorption and DFT that identify V$3d$–Sb$5p$ hybridization as an active driver of the CDW transition in \csvsb,\cite{Han2022AdvMat} and with resonant X‑ray scattering that reveals an Sb‑$5p$–assisted $2{\times}2{\times}2$ component conjoined with the kagome‑plane $2{\times}2{\times}1$ order in the three‑dimensional CDW state.\cite{Li2022NatCommun} Meanwhile, ARPES and quantum-oscillation studies show that the in-plane $2{\times}2$ reconstruction predominantly reconstructs V-derived pockets, whereas the central Sb-$p_z$ pocket at $\Gamma$ is comparatively less affected,\cite{ortiz2021physrevx} consistent with symmetry lowering primarily within the V network, with Sb-$p$ states contributing to the three-dimensional character of the order.

\section{Conclusion}

Orientation-dependent ${}^{51}$V NMR on \csvsb, with \rbvsb\ as a reference, yields two findings. 
First, \csvsb\ exhibits an interlayer-modulated $3\mathbf q$ charge order whose local environments are consistent with a $2\times2\times4$ stacking, while \rbvsb\ shows the uniform TrH $2\times2\times2$ order; this provides a bulk, site-selective discriminator between these three-dimensional CDW microstructures. 
Second, at $T_{\mathrm{CO}}\!\approx\!65$~K and for $\mathbf H_0\!\parallel c$, the ${}^{51}$V spectra reveal two inequivalent EFG environments that appear and coexist over a finite interval, indicating a reduction of the average in-plane rotational symmetry within the charge-ordered state.

The two-manifold spectra further indicate in-plane phase segregation and, together with conserved total satellite intensity and nearly temperature-independent $\nu_Q$ within each manifold, establish the bulk first-order character at $T_{\mathrm{CO}}$. 
NMR does not determine the in-plane wave vector and we therefore refrain from assigning a microscopic nematic mechanism; the data are, however, compatible with a uniaxial component reported by surface probes.
DFT calculations based on synchrotron-refined 90~K and 11~K structures capture the evolution from a narrow to a clustered distribution of EFG local tensors, consistent with inequivalent V environments. 
These results provide a quantitative, site-selective benchmark for distinguishing $2\times2\times4$ from $2\times2\times2$ stacking in the 135 kagome family.\\

\begingroup
\setlength{\parskip}{0pt}
\begin{acknowledgments}We thank Stuart Brown, Riku Yamamoto, Tri Thanh Chau (University of California, Los Angeles), and Rong Cong (National High Magnetic Field Laboratory) for insightful discussions. NMR measurements were performed at the National High Magnetic Field Laboratory, which is supported by the National Science Foundation under Cooperative Agreement No.~DMR-2128556 and the State of Florida. This work was supported by the U.S.\ Department of Energy, Office of Science, under Award No.~DE-SC0025712.
\end{acknowledgments}

\bibliography{apssamp}

\end{document}